# Towards peptide strings pharmacology

**Razvan Tudor Radulescu**

Molecular Concepts Research (MCR)

Muenster, Germany

E-mail: ratura@gmx.net

Mottos:        *"Nothing is more practical than theory."*
(Walther Nernst)

*",....So I hope to leave you with the impression that a new dimension is opening up for biochemistry-that of the electrons, a dimension in which a molecule assumes a new importance as a quantum mechanical framework. Every single atom with its electronic profile acquires a new and subtle personality. Ideas on which we built, till the present, such as the mass law action, lose much of their importance, while the whole system, with its great subtlety, begins, if I may say so, to smell of life. I also hope to have convinced you that the distance between those abstruse quantum mechanical calculations and the patient's bed is very small, and that further work on these lines may not only deepen our understanding but may also help to complete the armory of medicine with which we try to eliminate human suffering."*
(Albert Szent-Györgyi, New Biological Dimensions, Perspectives in Biology and Medicine, Summer 1961)

**ABSTRACT:** Peptide strings have been developed as a concept for the past fourteen years. They are proposed to basically entail various quantum states engendered by the physical interactions of proteins containing peptide portions that are similar to one another and, moreover, amino acid sequences that are sterically complementary to each other. In this survey, additional insights are presented that support the notion that peptide strings are likely a biophysical phenomenon that warrants further investigations. Specifically, these putative peptide strings traits are the capacity for wave-like interferences and electric current-like properties. Therefore, future experimental validation and quantification of these predicted features as well as application of this potential energy that is stored in the distinct shapes of proteins and peptides may prove beneficial in addressing challenges in physics and medicine. Within the latter field, the key discipline of pharmacology should be most fundamentally advanced by the peptide strings approach.

Ever since Hans Christian Ørsted, André Marie Ampère and Dominique François Arago originated the concept of electromagnetism that was subsequently further elaborated by Michael Faraday, James Clerk Maxwell and Heinrich Hertz, the underlying theoretical construct of an electromagnetic field has remained central to physics.

Within the same framework, it has recently been reported that modulating the transmembrane potential of various cells, including human melanocytes, -more precisely speaking, depolarizing these cells- can affect their shape and behavior, thereby inducing through epigenetic mechanisms a neoplastic-like phenotype at a distance (Blackiston et al., 2011). More recently, it has been shown that bacterial cells can communicate at a distance through long-range electrical signaling (Humphries et al., 2017).

Yet, an emerging question is whether, besides these so-called bioelectric signals, there may also exist another field that emanates from specifically biological, subcellular processes and, as such, be consistent both with Max Delbrück´s prediction of another law in physics intrinsic to biology as well as with Albert Szent-Györgyi´s prescient concepts on quantum biochemistry (Szent-Györgyi 1941), bioenergetics (Szent-Györgyi 1956) and bioelectronics (Szent-Györgyi 1968).

A potential hint to such biological/biochemical law could be the observation according to which porphyrins which are important metabolic compounds have the ability to self-assemble and give rise to electronic conductors owing to their (shape) complementariness at the molecular level (Drain 2002). A further indication along the same lines stems from measurements which suggest that the same type of molecules possesses wave-like properties (Hackermüller et al., 2003).

Independent of these explorations on the biophysical properties of porphyrins, I have deduced comparable putative features of peptides that I have collectively coined as "peptide strings" and which include wave-like properties for distinct peptides (Radulescu 2009). As a result, (allosteric) proteins with peptide portions which are similar to one another, yet also sterically (self)complementary to other peptide domains *resonate* with one another (Radulescu 2009), thus resembling other (less complex) chemical structures that have been predicted to display the characteristic of quantum mechanical resonance by virtue of their structural equivalence (Pauling 1932) and, moreover, paralleling aromatic amino acids for which resonance transfer of excitation energy has been shown (Karreman et al., 1958).

Similar to the above-mentioned bioelectric signals, peptide strings may equally constitute a possible epigenetic pathway for modulating cell fate, for instance during oncogenesis through a process I denominated "oncoprotein metastasis" (Radulescu 2007b; Radulescu 2008; Radulescu 2010; Radulescu 2013).

Along similar lines, it should also be noted that while Niels Jerne´s "mirror halls" of immunity (Jerne 1974) entailed mainly the reciprocal biological and biochemical relationship between antigens and the corresponding antigenic determinants on antibodies, peptide strings add a crucial dimension to peptide-based dynamic interactions which is the emergence of biophysical or, respectively, quantum biological states in living cells and organisms as a result of these interactions.

In the same context and consequently, such (post-translational) peptide strings waves would by definition also display the phenomenon of interference. Besides the purely physical parameter that might be amenable to a precise measurement, this would also entail the biological/biochemical feature of a *de facto* (enhanced) propagation or, respectively, neutralization of a given biological information encoded by a certain amino acid sequence (thus paralleling the pre-translational process of RNA interference).

For example, a subcellular predominance of proteins/peptides carrying (in an exposed manner) the (potentially oncogenic) amino acid sequence motif XLXCXE would predictably lead to an oncogenic state (or resonance) of the cells harboring such predominant molecules corresponding to an enhancement of the respective (oncogenic) wave (Fig. 1a).

On the other hand, a predominance of proteins/peptides carrying (in an exposed manner) an amino acid sequence that is functionally antagonistic to such XLXCXE motif due to a likely shape complementariness between these two different structures, such as the LFYKKV hexapeptide, would conceivably lead to an anti-oncogenic (quantum) state and thus resonance of the cells harboring such predominant molecules coinciding with an attenuation or even complete neutralization of the respective (oncogenic) wave (Fig. 1b).

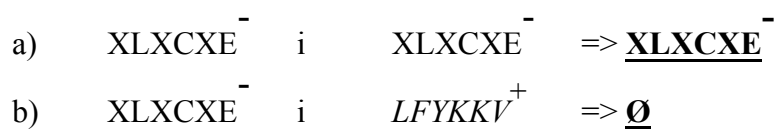

a)      XLXCXE⁻      i      XLXCXE⁻      => **XLXCXE⁻**

b)      XLXCXE⁻      i      *LFYKKV⁺*      => **Ø**

Fig. 1   Equivalents of wave interferences of peptide strings at the level of structurally and functionally similar (a) or antagonistic (b) amino acid sequences, yielding either an enhancement (a) or attenuation (b) of the biologic/biochemical information encoded by such sequences; „i" = interferes with.

In this context, tumor suppressor-derived peptide therapeutics may **resonate** with and thus activate *endogenous* tumor suppressor proteins, thus ultimately generating highly effective antineoplastic quantum states and fields across cells (Radulescu 2010), predictably at the expense of only small adverse effects (if any). This structural and functional ***equivalence*** (or, respectively, ***transcellular molecular harmony***) between a given (exogenously administered) therapeutic substance and endogenous molecules may also lead to a ***novel type of (significantly increased/improved) bioavailability*** compared to treatments that do not display such equivalence. For instance, the retinoblastoma tumor suppressor protein, briefly: RB (Radulescu 2016) and RB-derived anti-cancer peptides (Radulescu 2014a) may be regarded as ***functional isoforms of the same quantum state of a growth-inhibitory RB activity*** and, as such, may synergize with one another if they were brought into close subcellular proximity (facilitating physical interactions) and resonance. This could also serve as an important example for a fundamentally novel type of ***quantum pharmacology***. Within this new field that encompasses bionic or, respectively, biomimetic features, as previously ascertained (Radulescu 2014b), it would not be a given protein or peptide which is the main focus of diagnostics and therapeutics, but their common (abstract) function or, respectively, shared quantum state that would represent the essence.

Since, by contrast, classical chemotherapeutic substances and more modern growth factor receptor inhibitors do not have natural counterparts at the cellular level, they are likely not to induce resonance phenomena.

Furthermore, it is also interesting to note that the above-specified first insights on potential interference phenomena of peptide (energy) waves are reminiscent of the initial concepts on interference of light (energy) waves advanced and experimentally established by Dominique François Arago along with Augustin Jean Fresnel in the early 19th century.

Perhaps even more importantly, the above-specified dual (i.e. intrinsic and reciprocal) shape complementariness of these peptides may also enable these "biological quanta" (Radulescu 2009) to develop electric current-like properties whereby one shape would be propagated, hence corresponding to the electrons of a conventional electric current. Moreover, the direction of such peptide strings current would originate from one shape, thus corresponding to the cathode of an electric voltage, and subsequently proceed towards its complementary shape, corresponding to the anode of an electric voltage (Fig. 2). Within the same framework, such electron-like peptide sequences (e.g. oncopeptides) could be pharmacologically antagonized by sterically complementary, positron-like peptide sequences (e.g. tumor suppressor peptides).

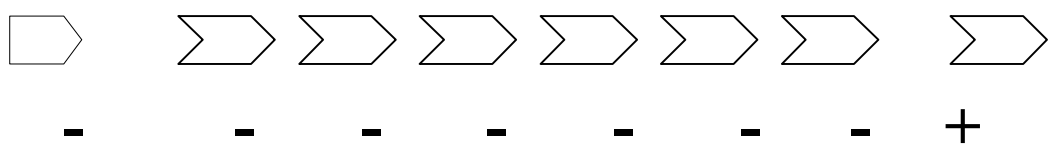

Fig. 2  Electric current-like properties of peptide strings constituted by peptides displaying shape complementariness both towards themselves and towards one another. This model is reminiscent of the structural and functional complementariness of electrons and their corresponding empty states (holes) in semiconductors.

It is evident that the implications of these considerations are far-reaching both for physics and biology. For instance, novel bioelectronic devices based on these envisaged peptide strings properties could be designed and tested, one of which might be a new type of semiconductors. Moreover, fundamentally novel medical treatments for a variety of diseases including cancer could equally be developed whereby pathologic subcellular states would be reversed in an epigenetic manner by peptide strings induced by specific therapeutic peptides.

It should also be noted that, intriguingly, the above-discussed peptide strings that are assumed to occur as a result of interactions between certain peptides (Radulescu 2009) fulfill Yoichiro Nambu´s definition of the physical mass as being a dynamical entity that comes into being through (molecular) interactions. As a consequence of the well-known mass-energy equivalence and, in addition, due to their wave-like properties (Radulescu 2009), peptide strings may therefore represent a novel form of (potential) energy and, as such, should be further investigated accordingly.

In analogy to the classical energy equations $E_{kin} = 1/2\ m\ v^2$ and $E = m\ c^2$, the above-outlined potential biophysical energy may be subject to an equation that introduces the ***localization*** *of a biomolecule (e.g. peptide or protein) at the subcellular level as a unit that intrinsically or, respectively, indirectly contains the concept of (information transfer) velocity* (Fig. 3). In other words, depending on where a distinct biomolecule (e.g. peptide or protein) is present (e.g. in the extracellular space *vs.* the cell nucleus), its capacity and speed of conveying a specific information to the cell differentiation and proliferation program is also different (e.g. relatively slow and redundant *vs.* relatively rapid and specific).

$$\mathbf{E}_f = m\ L_{seq}$$

Fig. 3          In this equation, $E_f$ is the biological or, respectively, biocybernetic energy that results from a given (sub)cellular function of a certain biomolecule (e.g. peptide or protein), $m$ is the mass of the biomolecule and $L_{seq}$ is the (sub)cellular localization sequence the latter of which could be quantified by establishing that the closer to the cell nucleus and the genes (thus, the greater its potential influence on gene regulatory processes and the fate of the cell) a biomolecule has located, the higher its $L_{seq}$ value.

One variation on this theme has already been illustrated in the context of the insulin-retinoblastoma tumor suppressor protein (RB) heterodimer formation whereby the $L_{seq}$ value has been defined in specific quantitative terms (Radulescu 2007a).

In this context, one additional analogy is noteworthy. As such, the number of intermediate molecules (and the cybernetic distance or, respectively, proximity) mediating the effect of a given biomolecule on (nuclear) DNA (Radulescu 2007a) is to some extent equivalent to the measure of electric resistance, as previously defined by Georg Simon Ohm. In other words, the less intermediates between a biomolecular cue and (its ultimate effect on) gene expression, the lower the (bio)cybernetic redundancy as well as resistance and thus the higher the (electric current´s intensity-like) cybernetic force and energy. In this context, the nucleocrine pathway (Radulescu 2015) might represent an example for such mimimally redundant, highly specific avenue for the transduction of biological information.

There is an interesting practical corollary to this insight. Accordingly, if such a biomolecular cue (with a need for a minimal number of intermediates for its ultimate action) is being targeted for therapeutic purposes, then predictably a lower resistance to therapy would ensue (Radulescu 2005a, Radulescu 2015).

Taken together, it may well turn out that, besides fractal features, as discovered by Benoît Mandelbrot, Nature may also comprise an intrinsic, at first sight hidden peptide strings-based (epigenetic) dimension whereby fractals and peptide strings may structurally and functionally intersect both at the microscopic and macroscopic level of biological organisms through quantum-like peptides and their induced geometric shapes and biophysical (quantum) states. In this context, it is worth noting that fractal structures, i.e. constructs showing self-similarity at various scales, have been detected in microglial cells (Karperien et al., 2013), nuclear chromatin (Metze 2013) and lung cancer (Lennon et al., 2015). Accordingly, it should be interesting to explore if peptide strings- which share with fractals the feature of self-similarity- are equally present in these cases.

This fundamental dimension may ultimately solve the so-called *paradox* that had been formulated by Max Delbrück several decades ago according to which biological phenomena obey physical laws and principles, yet we cannot understand them fully on the basis of such classical physics. In contrast to the genes which, as already revealed by the results of the Human Genome Project obtained almost 2 decades ago, cannot account neither for the differences between various species (e.g. between man and worms or plants whereby the differences in the respective numbers of genes were found to be rather minor) nor for various aspects of life including its pathogenetic processes (therefore calling into question the sense of therapeutically intended gene replacement strategies such as the CRISPR/Cas method), proteins and peptides may hold the key for biologic diversity, e.g. through epigenetic effects of peptide strings, as described here. This particular type of proteomic data should significantly contribute to the awaited complete understanding of cellular life through the novel insights that they provide and which result from combining physical, more precisely quantum-mechanical principles with distinct biochemical and biophysical properties of peptides. In this way, the protein and peptide research directions initiated by scientists such as Emil Fischer, Linus Pauling and Vincent du Vigneaud in the first half of the 20th century and put to some extent on hold by the discovery of the DNA double helix and further gene-related developments (Radulescu 2005b) would be fully accomplished with predictably profound benefits both for the biomedical field and physics.

***Note added in proof***: Consistent with the electric current-like properties of peptide strings described here, Ing *et al.* have recently (*J. Phys. Chem. B.* 122: 10403-10423, 2018) underscored in their paper on long-range conductivity in protein and peptide bioelectronic materials the "impressive electronic properties" of proteins and peptides which, however, according to these investigators, are "not readily understood". Therefore, the transmission of core information through sequential and coherent conformational changes of proteins intrinsic to peptide strings as well as their structurally binary character could indeed fill this crucial knowledge gap. At the same time, it should be emphasized that peptide strings are yet another example for the quantum physical wave-particle duality, i.e. these specific peptides that are capable of generating peptide strings comply with both the definition of, on the one hand, a particle or, respectively, substance and, on the other hand, the definition of a wave or, respectively, field, similar to other particles such as electrons or photons.